\documentclass[11pt, a4paper]{article}   	% use "amsart" instead of "article" for AMSLaTeX format
\usepackage{geometry}                		% See geometry.pdf to learn the layout options. There are lots.
\usepackage[parfill]{parskip}    		% Activate to begin paragraphs with an empty line rather than an indent
\usepackage{graphicx}				% Use pdf, png, jpg, or eps§ with pdflatex; use eps in DVI mode
\usepackage{color}
\usepackage{amssymb}
\usepackage{fleqn}
\usepackage[utf8]{inputenc}
\usepackage{hyperref}

\begin{document}

\title{{The natural parameterization of cosmic neutrino oscillations}}
\author{Andrea Palladino$^{1}$ and Francesco Vissani$^{1,2}$\\[2ex] \small
$^{1}$Gran Sasso Science Institute, INFN, L'Aquila (AQ), Italy\\ \small
$^{2}$Laboratori Nazionali del Gran Sasso, INFN, Assergi (AQ), Italy}
%\emailAdd{andrea.palladino@gssi.infn.it}
%\emailAdd{francesco.vissani@lngs.infn.it}

\date{}							% Activate to display a given date or no date
\maketitle

\begin{abstract}

The natural parameterization of vacuum oscillations in three neutrino flavors is studied. 
Compact and exact relations of its three  parameters with the ordinary three mixing angles and CP violating phase are obtained. 
Its usefulness is illustrated by considering various applications: 
the study of the flavor ratio and of its uncertainties, the comparison of expectations and observations 
in the flavor triangle, the intensity of the signal due to Glashow resonance.
The results in the literature are easily reproduced and in particular the 
recently obtained agreement of the observations of IceCube with the hypothesis of cosmic neutrino oscillations is confirmed.
It is argued that a Gaussian treatment of the errors appropriately describes the effects of the uncertainties on neutrino oscillation parameters. 
\end{abstract}

%\maketitle
%\flushbottom

%\renewcommand\contentsname{\vskip-4mm}{\tableofcontents}

%\section{}
%\subsection{}
\section{Introduction}
After IceCube results, see e.g.\ \cite{ice1,ice2,ice3}, the importance of a precise description of oscillations has increased greatly. In the present paper we discuss a natural, easy-to-use and completely general choice of the relevant parameters. We illustrate its usefulness by quantifying the impact of the uncertainties on various physical quantities,
implied by the imprecise knowledge on 
oscillations.

We begin by recalling the main achievements in the discussion of cosmic neutrino oscillations. 
The general formula for the vacuum averaged oscillations  was given in \cite{pontecorvo}. 
Ref.~\cite{jlp} studied for the first time the implications of the observed oscillation phenomena on cosmic neutrinos.  Various authors remarked the possibility to measure flavor ratios, possibly aiming to constrain the 
parameters of oscillations, e.g.\ \cite{m1,Barenboim,m2,m3}.  
The relevance of oscillations for the interpretation of 
Glashow resonance was noted  in \cite{piton}.
In \cite{cost} 
the single parameter that rules cosmic neutrino oscillations and 
 depends linearly upon unknown quantities was identified; then,
it was remarked \cite{vis} that this leads to a strong correlation between  the effect of the oscillation parameters 
on the probabilities of oscillation, lessening chances of measuring the oscillation parameters.
Non-linear effects were studied in various subsequent papers including \cite{nl1,nl2,nl3}. An interesting expression for all relevant parameters  of vacuum oscillations 
was proposed in \cite{nl45}  within 
a scheme of approximation aimed to improve the 
tribimaximal mixing matrix.   In \cite{nl45,nl5,nl4} 
an expansion in second order of small parameters of this  parameterization   was obtained and applied to the study  of cosmic neutrinos. 
The consistency of vacuum oscillations and  
IceCube observations was discussed in 
\cite{mena2}, 
\cite{Palladino:2015zua},
\cite{Aartsen:2015ivb} compare with
\cite{mena1}.
In \cite{Palladino:2015zua} the impact on the flavor ratio of the uncertainties on oscillation parameters was analyzed 
and the present work  develops the discussion.

In this work, we show that the parameterization of neutrino oscillations in vacuum introduced in~\cite{nl45} 
can be promoted to an exact parameterization 
and can be argued to be the natural parameterization for the discussion of  oscillations of cosmic neutrinos.\footnote{An alternative choice of the parameterization that gives different insight on the allowed ranges of the oscillation probabilities is discussed in~\cite{Xu:2014via}.} We obtain new and exact expressions of the 
three parameters in terms of the known mixing angles and CP violating phase.
We evaluate their numerical values and uncertainties, illustrating their usefulness by discussing three applications: 
1)~we compare the predicted flavor fractions and those that are allowed by the present observations; 
2)~we quantify the uncertainties in the prediction of the fraction of muon neutrinos due to oscillations; 
3)~we argue that,  
even after accounting very conservatively for the uncertainties of oscillations, 
the intensity of the Glashow resonance \cite{glash0} differs greatly 
in the alternative cases of $pp$- and $p\gamma$-production 
as remarked in \cite{piton} and later discussed in  \cite{nl5,glash,glash1,glash2}.
Throughout this work, we argue that a Gaussian treatment of the errors of these natural parameters is quite adequate for the present precision.

 \section{Natural parameters for three flavor vacuum oscillations}

 In this section, we motivate, define and analyze a parameterization of vacuum neutrino oscillations, elucidating the relationship between this and other parameterizations.

We begin by counting the number of  independent vacuum oscillation parameters~\cite{pontecorvo} 
\begin{equation}
P_{\ell\ell'}=\sum_{i=1}^n |U_{\ell i}^2| |U_{\ell' i}^2|
\mbox{ where $\ell=e,\mu,\tau...$}
\end{equation}
 in the case of  $n$ light neutrinos.
 
The vacuum oscillation formula depends upon the 
squares of the leptonic mixing matrix $|U_{\ell i}^2|$.
They correspond to $(n-1)^2$ independent parameters, as it is clear considering  
all  $|U_{\ell i}^2|$ as independent (albeit constrained)  parameters
except the ones of the first row and column, that can be obtained from unitarity,  e.g., $ |U_{\mu 1}^2|=1-\sum_{i=2}^n |U_{\mu i}^2|$. 

But the vacuum oscillation probabilities 
$P_{\ell\ell'}$ are also {\em symmetric} in the exchange of the flavor indices
$\ell\leftrightarrow \ell'$, thus they require less parameters. Since we can again rely on similar unitarity relations when we sum on all flavors, namely 
$ \sum_\ell P_{\ell\ell'}=1$, this implies that the number of independent parameter is just 
$n (n-1)/2$.  This means that when $n=3$ we have 3 independent parameters, when $n=4$ we have 6 of them, etc. 

From here on and in view of the present experimental situation,  we focus on the three flavor case ($n=3$) where we have three  parameters, as first remarked (to the best of our knowledge) in ref.~\cite{nl45}.
%after Eq.\hskip0.5ex(20) there.}

 \subsection{Motivations and definition \label{motta}}

Next, we motivate and introduce the choice of the three natural parameters. 
%Let us define now the three relevant parameters 
%and describe their relation with the four standard mixing angles and CP violating phase, 
%$\theta_{12}, \theta_{23}, \theta_{13}, \delta$. 
The parameters $P_0,P_1,P_2$ are defined as follow,
\begin{equation}\label{our}
 P_0=\frac{P_{ee}-\frac{1}{3}}{2}  \ , \
P_1  =\frac{P_{e \mu}-P_{e \tau}}{2}  \ , \ 
 P_2 =\frac{P_{\mu \mu}+P_{\tau \tau}-2P_{\mu \tau}}{4}
\end{equation}
We can write in terms of $P_0,P_1,P_2$ 
the matrix that contains the probabilities of oscillations of cosmic neutrinos. This is the following symmetric matrix, 
\begin{equation} \label{pupp}
\mathcal{P}=\left( \begin{array}{ccc}
\frac{1}{3}+2 P_0 & \frac{1}{3}-P_0+P_1 & \frac{1}{3}-P_0-P_1 \\
 & \frac{1}{3}+\frac{P_0}{2}-P_1+P_2 & \frac{1}{3}+\frac{P_0}{2}-P_2 \\
 &  & \frac{1}{3}+\frac{P_0}{2}+P_1+P_2 \end{array} \right) 
 \end{equation} 
It  acts on the vector of fluxes before oscillations   $F^0=(F_e^0,F_\mu^0,F_\tau^0)$ just as $F=\mathcal{P}\ F^0$, giving 
the vector of fluxes observed after oscillations, $F=(F_e,F_\mu,F_\tau)$.

We would like to give our 
reasons (that are 
largely based on the available experimental information) 
why we consider  that Eq.~\ref{our}  is the optimal choice of parameters.
\begin{enumerate}
\item The oscillation probability that  is singled out in $P_0$ is 
$P_{ee}$, that is well-known (being directly measured by 
low energy solar neutrino experiments and probed also by 
reactor and high energy solar neutrino experiments).
\item The difference of $P_{e \mu}-P_{e \tau}$ contains  most of the uncertainties.
\item The last combination of oscillations probabilities, $P_2$, 
is positive and very small.
\item A specific choice of the overall coefficients 
is adopted in order to have coefficients that are either zero or close to 1 in the expressions of all oscillation 
probabilities,  Eq.~\ref{pupp}.
\item Setting $P_0=P_1=P_2=0$, all
oscillation probabilities become $P_{\ell\ell'}=1/3$, namely, any information on the original flavor is lost:
%see again Eq.~\ref{pupp}. 
The three parameters describe the potentially measurable
information on flavor that survives cosmic neutrino 
oscillations. 
%and that can be potentially measured.  
\end{enumerate}
The above argument clarifies   
that it is possible to introduce such a parameterization directly, 
 without the need to associate it to a specific  scheme of approximation of the mixing matrix, but rather, keeping it exact. 
 However, various approximation schemes used in the literature 
 have allowed to uncover the most interesting properties: The second one is known since \cite{cost} while the first evidences of the third one  were found in \cite{nl45}.
%Other choices of parameters, that are equivalent to the one presented here,  were  proposed in \cite{nl45} 
%and \cite{nl5} to summarize the results of an approximate description of the parameters of oscillations. 
%See Eq.\hskip0.5ex(19) of
%\cite{nl45}  and Eq.\hskip0.5ex(2.5) of \cite{nl5}, noting respectively the symbols $\mathcal{O}(\epsilon^3)$ in \cite{nl45}  and the 
%approximate equality symbol $\simeq$ that have been used in \cite{nl5}. 
A detailed comparison with other parameterizations used in the literature is offered in Sect~\ref{xxx}.

 \subsection{Connection with the standard parameters of neutrino mixing\label{cx2}}

Compact and useful expressions of the natural parameters
in terms of four standard parameters, the 
mixing angles $\theta_{12}, \theta_{23}, \theta_{13}$  
and CP phase violation phase $\delta$, 
are as follows,
\begin{equation}
\begin{array}{l}\displaystyle
P_0 = \frac{1}{2} \bigg\{ (1-\epsilon^2) \bigg[ 1- \frac{\sin^2 \theta_{12}}{2} \bigg] + \epsilon^2-\frac{1}{3} \bigg\}  \\[3ex]
\displaystyle
P_1 = \frac{1-\epsilon}{2} \bigg\{ \gamma \cos 2\theta_{12}+\beta \frac{1-3 \epsilon}{2} \bigg\} \\[3ex]
\displaystyle
P_2 = \frac{1}{2} \bigg\{ \gamma^2+\frac{3}{4} \beta^2 (1-\epsilon)^2 \bigg\} 
\end{array}
\end{equation}
where we introduce for convenience 
the following 4 small parameters,
%\begin{equation}
%\begin{array}{l}
%\epsilon=\sin^2 \theta_{13} \nonumber \\
%\alpha=\sin \theta_{13} \cos \delta \sin 2 \theta_{12} \sin 2 \theta_{23}    \\
%\beta=\cos 2 \theta_{23}   \nonumber \\
%\gamma=\alpha-\frac{\beta}{2} \cos 2\theta_{12} (1+\epsilon) 
%\end{array}
%\end{equation}
\begin{equation}
\begin{array}{ll}
\epsilon=\sin^2 \theta_{13} &
\alpha=\sin \theta_{13} \cos \delta \sin 2 \theta_{12} \sin 2 \theta_{23}    \\[2ex]
\beta=\cos 2 \theta_{23}   &
\gamma=\alpha-\frac{\beta}{2} \cos 2\theta_{12} (1+\epsilon) 
\end{array}
\end{equation}
These expressions are new and exact. Note property 3 
listed in Sect.~\ref{motta} of this parameterization.

\begin{table}[t]
\centering
\begin{tabular}{|c|c|c|}
\hline
Parameter & Mean value & Standard deviation \\
\hline
$P_0$ & 0.109 & 0.005 \\
\hline
$P_1$ & 0.000 & 0.029 \\
\hline
$P_2$ & 0.010 & 0.007 \\
\hline
\end{tabular}
\caption{\em \small Table of present values and errors of the natural parameters.}
\label{tab1}
\end{table}

 \subsection{Numerical analysis\label{qa}}

The parameters $\alpha,\beta,\gamma$ are small and 
to date not known precisely, whereas   $\epsilon$ is very small and precisely known.
We can then order these parameters according to their (presumed)  size, and  
consider $\sin^22\theta_{12}$ as zeroth order; $\sin\theta_{13}$, $\cos2 \theta_{23}$, $\alpha,\beta,\gamma$ of first order; 
$\epsilon$ of second order.
In the same sense, 
$P_0$ is a zeroth-order parameter; $P_1$ is first order in $\alpha$ and $\beta$;  
%and agrees with the leading order expression given in \cite{cost}; 
 $P_2$ is second order in $\alpha$ and $\beta$. Note that $P_2$ is bound to be positive.
%\comment{It should be remarked that, till here,
%we are still discussing the exact probabilities of oscillation without introducing 
%any approximation; the whole point is to identify the natural parameters 
%for their description.}

\begin{figure}[t]
\centering
\includegraphics[scale=0.3]{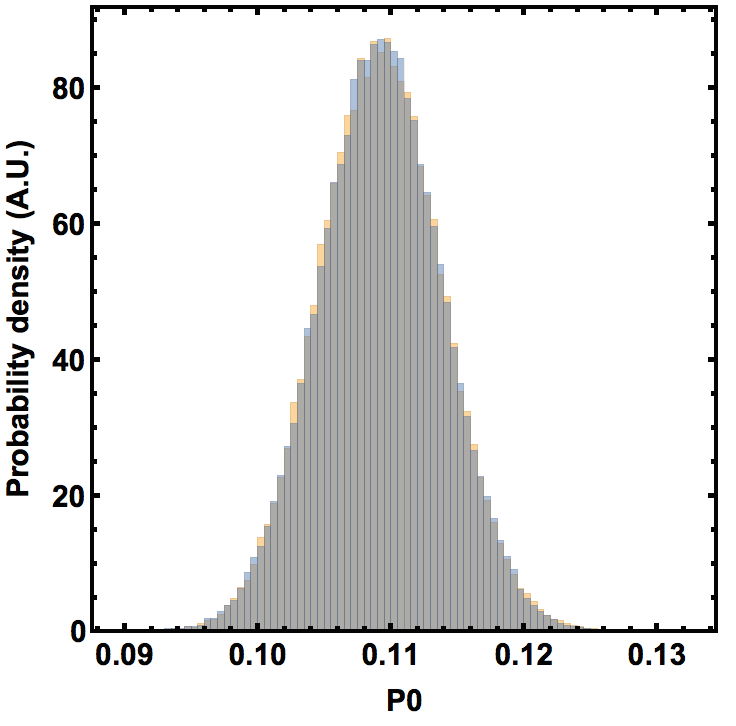}
\includegraphics[scale=0.3]{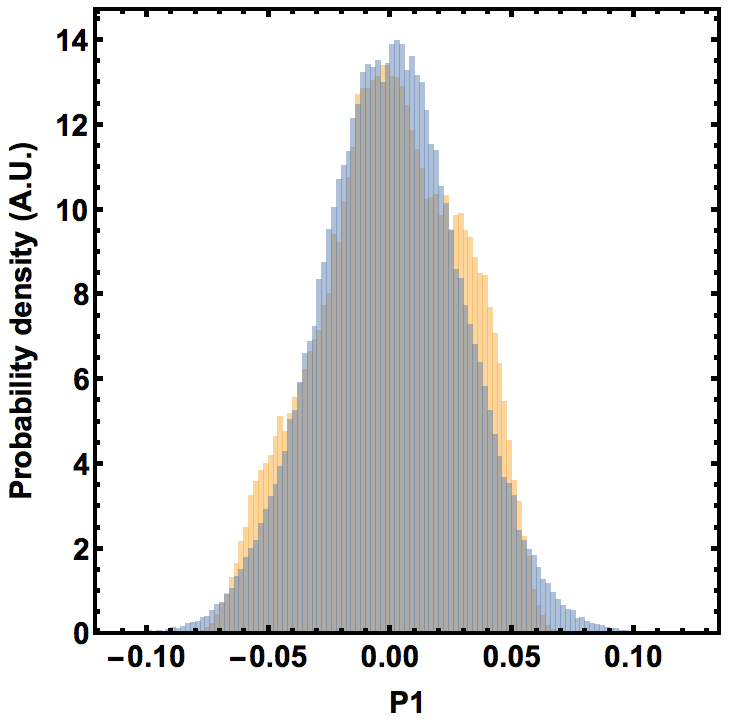}
\includegraphics[scale=0.31]{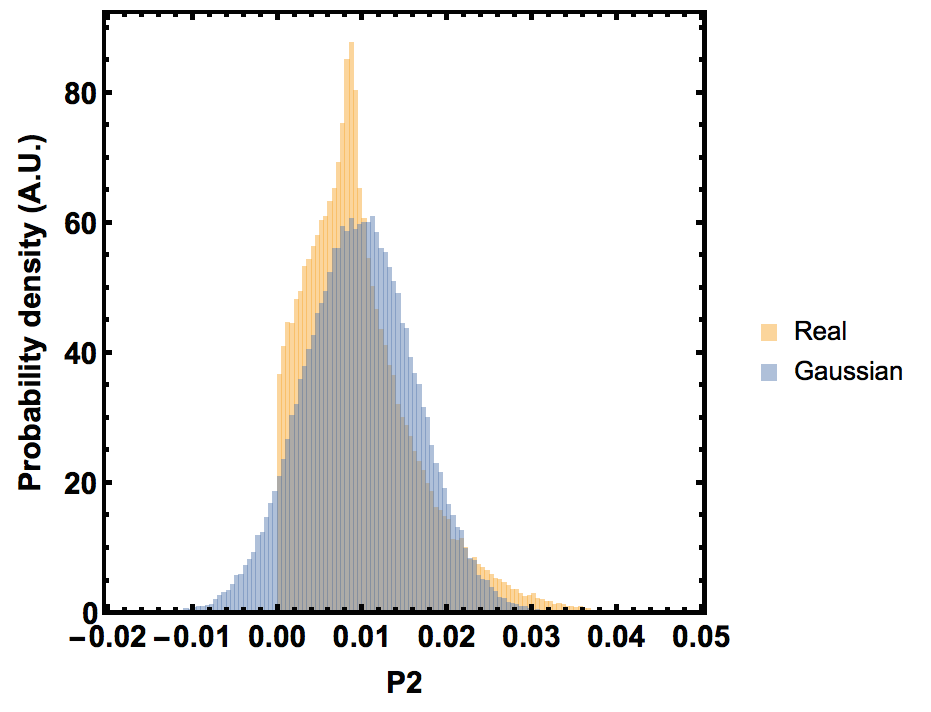}
\caption{\em \small Distribution of the natural parameters $P_0$, $P_1$ and $P_2$, due to the uncertainties in the mixing angles and the phase of leptonic CP violation.}
\label{pardist}
\end{figure}

Using the present knowledge of mixing angle and CP 
violating phase \cite{Gonzalez-Garcia:2014bfa}, we obtain the values and the errors of the natural parameters. We show the results in the Table~\ref{tab1}, assuming normal mass hierarchy. It is easy to repeat the same steps with inverted hierarchy, but the differences are not large. 
 From this Table we notice that with present data  the average values obey  
 $\langle P_0 \rangle \gg 
 \langle P_1 \rangle\simeq \langle P_2\rangle $ whereas their variances obey $\delta P_1 \gg \delta P_0 \simeq \delta P_2$.  
$P_0$ is well known, because is related to survival probability of solar low energy neutrinos and $\theta_{13}$ or $\epsilon$ is well measured by reactor experiments.  As we see from figure \ref{pardist}, 
$P_0$ and $P_1$ are well represented by Gaussian functions; $P_2$ is not Gaussian but it is a very small parameter. For these reasons, as we argue in the rest of this work, we can use a Gaussian approximation without introducing severe inaccuracies in the numerical analysis of the oscillations. 
This is a new result, that allows one to 
obtain convenient analytical expressions for different examined quantities
and to quantify easily the uncertainties.

The probabilities of oscillation given in Eq.~\ref{pupp} have a very simple form: they depend linearly upon the natural parameters. Moreover, in first approximation, they could be expressed only in terms of $P_0$, because $P_1$ and $P_2$ give small corrections. Using the value of Table \ref{tab1} and the natural parameterization of oscillation matrix, we obtain the probabilities of oscillations, % for cosmic neutrinos, 
\begin{equation}
\begin{array}{ll}
P_{ee}=0.552 \pm 0.010,   & P_{e \mu} = P_{e \tau} =0.224 \pm 0.029 \nonumber \\ 
P_{\mu \tau}=0.378 \pm 0.008, & 
P_{\mu \mu}=P_{\tau \tau} =0.398 \pm 0.029 
\end{array}
\end{equation}
Two couples of probabilities have (almost) the same values, because with the present best fit value  $\langle P_1 \rangle=0$ and the numerical differences between these expressions are small.

 \subsection{Comparison with other parameterizations\label{xxx}}

First of all, we consider the leading order in the small parameters $\theta_{13}$ and 
$\cos2\theta_{23}$.  
The parameter $y$ that was introduced in 
 \cite{cost}, namely 
\begin{equation}
y=\frac{1}{4}\sin^22\theta_{12}\; \cos2\theta_{23}+
\frac{1}{2} \sin 2\theta_{12}\; \cos 2\theta_{12}\; \theta_{13}\; \cos\delta
\end{equation}
coincides at this order with $P_1$, while   
higher order terms are neglected. 
In  \cite{cost}, the errors on $P_{e\mu}$ and $P_{\mu\mu}$ were estimated to be 0.05;  after 10 years, these errors  amount to 0.03 and we see from table \ref{tab1} that, still, the uncertainty   
is mostly due to $P_1$. 
Equivalent parameters have been used 
also by other authors: E.g., in  \cite{nl45},
we read that  
\begin{quote}
{\em A ``universal'' parameter related to B has been noted previously in the literature. It is called $-\Delta$ in Z.\ Z.\ Xing, Phys.\ Rev.\ D 74, 013009 (2006), and $+\Delta$ in W.\ Rodejohann, JCAP 0701, 029 (2007).}
\end{quote}
although, curiously, reference \cite{cost} is not mentioned. 
Note that the parameter $y$ or equivalently $\Delta$ satisfies  
properties 2 and 4 of Sect.~\ref{motta}.

 As already mentioned, a  three-parameters  
 description of the probabilities of 
 vacuum oscillations with
 was first introduced in  
   \cite{nl45}, where 
   the parameters $A,B,C$  were defined:
   see Eq.\hskip0.5ex(19) and 
   note the symbol $\mathcal{O}(\epsilon^3)$ used there   
    to  emphasize the use of quadratic expression in  the 
small parameters  that quantify the deviation from tribimaximal mixing, collectively denoted as $\epsilon$. 
This approximation is even better than the linear approximation, and thus was argued to be 
 adequate for the present needs  \cite{nl45}. However, there are two evident shortcomings in
the type of procedure adopted there to introduce 
the new parameterization:  
The emphasis on tribimaximal mixing 
given in Eq.\hskip0.5ex(7) and see Eq.\hskip0.5ex(16) of  reference 
\cite{nl45} 
is felt as artificial to date, especially now 
that  the measurements
showed that $\theta_{13}$ is non-zero, 
contradicting the most interesting prediction of 
tribimaximal mixing. Moreover,  the efforts used to obtain a quadratic the expansion is also unnecessary: the 
parameterization can  be promoted without significant efforts to an exact one, as the expressions in Sect.~\ref{cx2} are  valid to all orders in $\epsilon$ and easy to use.
Thus, a direct, valid and advantageous procedure is to
introduce  the parameterization since the start, as in Sect.~\ref{motta}.
   The detailed relation with our parameterization is,
\begin{equation}
A = 9\, P_0 - 1,\  B = 18\, P_1 \mbox{ and } C = 18\, P_2.
\end{equation}
Both of them share features 1, 2, 3 discussed in  
Sect.~\ref{motta} and they  can be termed as 
natural. In the following we adopt $P_0, P_1, P_2$ due to features 4 and 5 of Sect.~\ref{motta} and to the fact that symbols reflect the hierarchy noted in  Sect.~\ref{qa}.
The parameterizations are however equivalent and it is 
easy to compare the results obtained with them.

Another equivalent parameterization  
was used in \cite{nl5}. This begins from the ``universal'' parameter of 
\cite{cost} and improves the description of the oscillation 
probabilities by introducing a new parameter.
Also this parameterization is introduced in connection to
tribimaximal mixing and using the same quadratic expansion of the previous parameterization: 
note the  symbol $\simeq$
of approximate equality in 
 Eq.\hskip0.5ex(2.5) of \cite{nl5}. Thus, the same comments 
 on the methodology  made just above  apply also to this case.
The relation with our parameterization is, 
\begin{equation}
\Delta=  P_1 \mbox{ and } \frac{\overline{\Delta}^2}{2} = P_2
\end{equation}
A third parameter is not introduced, being replaced by   
$\theta_{12}$ and $|U_{e3}|=\sin\theta_{13}$. Properties
2, 3 and 4 of Sect.~\ref{motta} are all satisfied. 
Therefore,  assumptions and results can be compared easily: e.g., Tab.~\ref{tab1} implies $P_2<0.017$ (resp., 0.031)
at 1 sigma (resp., 3 sigma) whereas the value 
given in Eq.~16 of  \cite{nl5} 
implies $P_2<0.0145$ (resp., 0.0465).
Similarly, 
	the values $\Delta \simeq 0.02$
	and $\overline{\Delta}^2\simeq 0.008$ quoted there correspond to $P_1\simeq 0.02$ and $P_2\simeq 0.004$,
	that are included in the 
	1 sigma range of Tab.~\ref{tab1}.
The differences are due to the improved  measurements  of the oscillation parameters since 2007 and in particular to the inclusion of $\theta_{13}$ that now is measured and known to be non-zero.\footnote{Let us repeat that this differs from what was expected from the  tribimaximal mixing scheme, that has been emphasized  in \cite{nl45} and \cite{nl5}.}
Note finally that in Fig.~1 of  \cite{nl45} the expected ranges of the parameters are presented, and these can be 
compared with our Fig.~\ref{pardist} and Tab.~\ref{tab1}.

In the literature
also other different linear combinations of the parameters have been considered, see e.g.~\cite{Xu:2014via}. The parameters $X,Y,Z$ introduced there,  however, do not satisfy features 1, 2, 3 and 5 of Sect.~\ref{motta}, and in this technical and restricted sense, we do not call them `natural'. 
The choice of  parameters in \cite{Xu:2014via} 
has its own motivations but it is less useful to keep under control the impact of the uncertainties on oscillations that is one of the main
goal of the rest of this work.

\section{Applications\label{appli}}

We will consider two quantities, that are affected by oscillations; the flavor ratios and the fraction of events due to Glashow resonance. 

We denote the fractions 
of $\nu_\ell$ at source and the one at Earth
(i.e., after oscillations)  respectively as,
\begin{equation}
\xi_\ell^0=F_\ell^0/\sum_{\ell} F_\ell^0
\mbox{ and }
\xi_\ell=F_\ell/\sum_{\ell} F_\ell
\end{equation} 
%and the fraction of $\nu_\ell$  with $l$,
where of course $\sum_{\ell} F_\ell^0=\sum_{\ell} F_\ell$. 
Suppose that the initial flavor ratio is given by 
\begin{equation}
(\xi_e^0,\xi_\mu^0,
\xi_\tau^0)=(1-g-h,g,h). 
\end{equation}
After propagation the flavor ratio is modified as follows, 
%\begin{eqnarray}
%\xi_e &=&\frac{1}{3}+P_0(2-3g-3h)+P_1(g-h) \nonumber \\
%\xi_\mu&=&\frac{1}{3}-\frac{P_0}{2}(2-3g-3h)+P_1(1-2g-h)+P_2 (g-h) \nonumber \\
%\xi_\tau&=&\frac{1}{3}-\frac{P_0}{2}(2-3g-3h)-P_1(1-g-2h)-P_2 (g-h) \nonumber 
%\end{eqnarray}
%$$
\begin{equation}
\begin{array}{l}
\xi_e=\frac{1}{3} + (2 - 3 g - 3 h) P_0 + (g-h) P_1 \label{frace} \\[2ex]
\xi_\mu= \frac{1}{3}  + \frac{1}{2}  (-2 + 3 g + 3 h) P_0 + (1 - 2 g - h) P_1 +  (g - h) P_2  \label{fracmu} \\[2ex]  
\xi_\tau = \frac{1}{3} + \frac{1}{2}  (-2 + 3 g + 3 h) P_0 + (-1 + g + 2 h) P_1 -  (g-h) P_2  \label{fractau}
\end{array}
\end{equation}
%$$
Below, we will emphasize $\xi_\mu$ since it is quite directly connected to an observable quantity, namely, the fraction of track-type events.

\begin{figure}[t]
\centering
\includegraphics[scale=0.6]{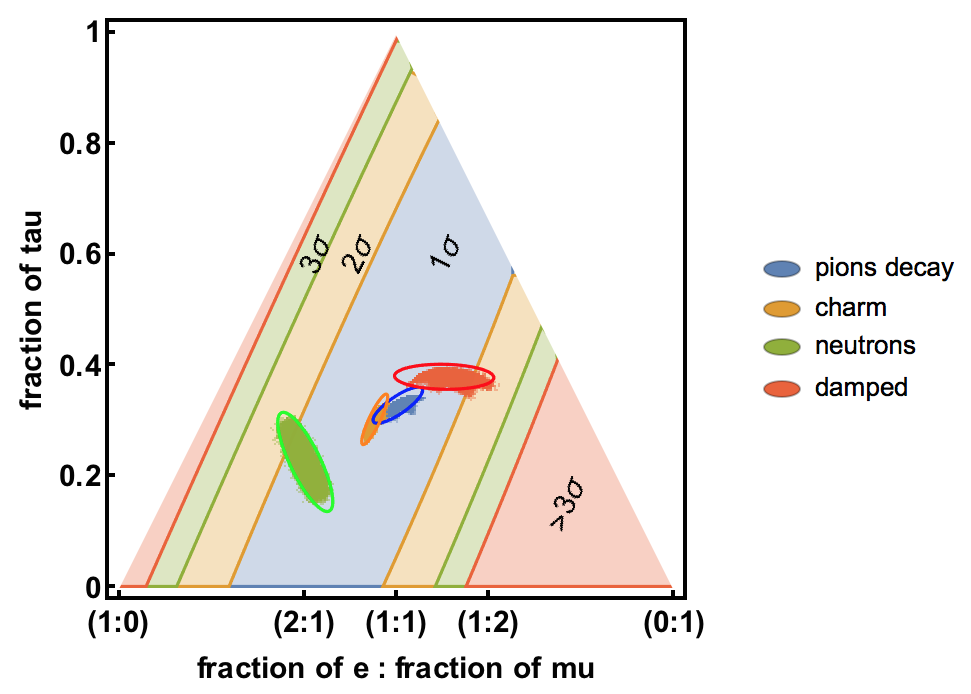}
\caption{\em \small Flavor triangle. 
The present observational information on the 
flavor composition of cosmic neutrinos at 1, 2 and 3$\sigma$  analyzed assuming $\alpha=2$ 
is compared with the 
expectations derived for four different hypotheses on the mechanism of production of the neutrinos. The ellipses derive from a Gaussian treatment of the 
error based on Table~\ref{tab1}, see also Appendix~\ref{aa}.}
\label{flavour}
\end{figure}

\subsection{Flavor ratio after the oscillations}
A first application of the natural parameterization is the study of the flavor ratio of neutrinos, considering different mechanisms of production.\footnote{Intermediate possibilities have been also considered in \cite{glash,cdg}.} 
The impact of uncertainties on the mixing angles 
and CP violating phase on the triangle of the flavors was first discussed in \cite{Barenboim}. Here we will update the analysis by using  updated values of the uncertainties. 
We will  verify that a Gaussian treatment of the natural parameters leads to results in 
good agreement with Monte Carlo simulation and, moreover, 
we will compare our theoretical analysis with three years IceCube HESE data + passing muons. 

\begin{figure}[t]
\centering
{\color{white}...}\includegraphics[scale=0.6]{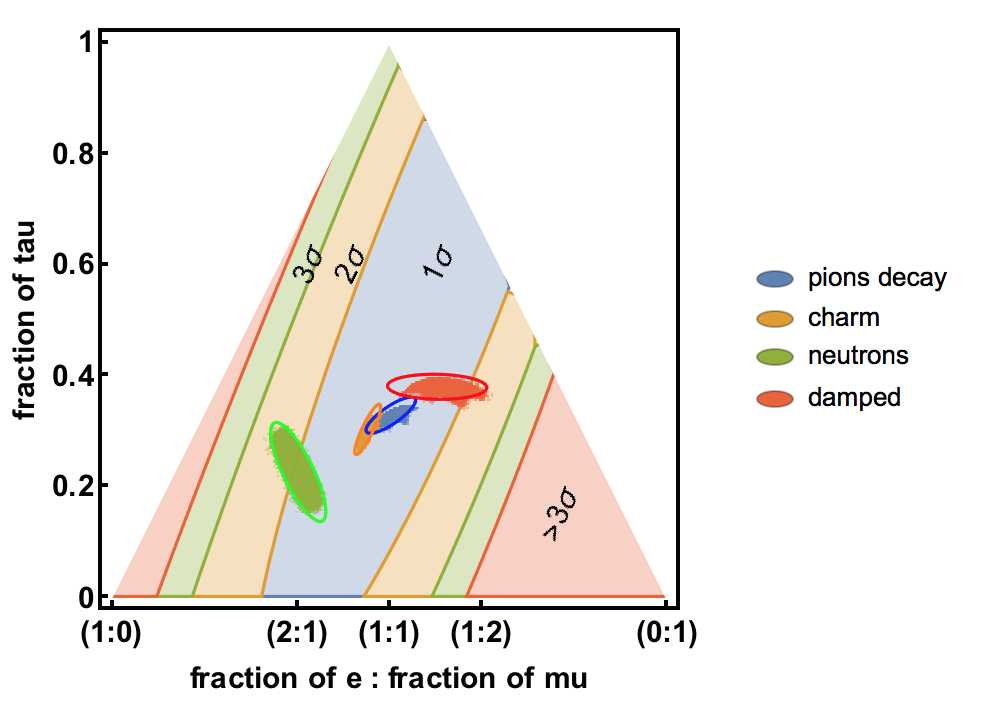}
\caption{\em \small Same as previous figure but using the spectral index $\alpha=2.3$ in the analysis of the IceCube data.}
\label{flavour2}
\end{figure}

We consider: \\
1. pions decay ($h=0,g=2/3$; blue); \\
2. neutrons decay ($h=0,g=0$; green); \\
3. damped muons ($h=0,g=1$; red); \\
4. charm mesons ($h=0,g=1/2$; orange). \\
Using Eqs.~(\ref{fracmu}), we represent the allowed regions by propagating the errors on the predictions by a 
Monte Carlo simulation; this gives the 4 dotted regions of  the flavor triangles in  
Fig.~\ref{flavour} and \ref{flavour2}.

These regions  can be compared with those obtained with a Gaussian treatment of the errors on $P_0$, $P_1$ and $P_2$. Following 
the implementation of Appendix~\ref{aa}, 
we obtain the 4 ellipses of  Fig.~\ref{flavour} and \ref{flavour2} 
that enclose the 99\% CL regions. 
We see that the differences between these the Montecarlo and the Gaussian 
treatments are not very important. The Gaussian approach
seems to be appropriate for the present needs. Note that the latter is  
significantly
easier to implement.

Let us repeat that the expected theoretical regions (the dotted areas obtained by Montecarlo and 
the elliptic curves  corresponding to Gaussian approximation) depend only by initial flavor ratio and are not affected by the energy spectrum of the neutrinos, that is assumed to be universal for all neutrinos. 
In other words, the four 
theoretical regions shown on the flavor triangles of Fig.~\ref{flavour}  are just identical
to those of Fig.~\ref{flavour2}.

On the contrary, the 1$\sigma$, 2$\sigma$ and 3$\sigma$ zones do depend on the 
energy distribution of the neutrinos. The confidence levels, indicated in the flavor triangle, correspond to the result of the data analysis of IceCube events (the high energy starting events, whose initial vertex is in the detectors and the passing muons, i.e.\ the thrugoing muons)  discussed in  \cite{Aartsen:2015ivb} and \cite{Palladino:2015zua}. They have been obtained  assuming a power law distribution given by,
%The theoretical elliptic curves on the triangle, of figure \ref{flavour}, depend only by initial flavour ratio and are not influenced by energy spectrum of neutrinos. On the other side, the 1$\sigma$, 2$\sigma$ and 3$\sigma$ zones are influenced by energy distribution of neutrinos. These confidence levels, indicated in the triangle, correspond to the result of the data analysis of IceCube events (HESE and passing muons), as discussed in  \cite{Aartsen:2015ivb}\cite{Palladino:2015zua}, considering a power law distribution given by:
\begin{equation}
\frac{d \phi}{d E_\nu}= \phi_0 E_\nu^{-\alpha}
\end{equation}
Thus, the observed flavor ratio is a function of spectral index $\alpha$. 
In Fig.~\ref{flavour} we have used the value preferred by the simplest theoretical expectations, namely 
$\alpha=2$. 

For comparison, we have shown also the case $\alpha=2.3$ in Fig.~\ref{flavour2}, namely the best fit value of the  dataset of high energy starting events collected by IceCube in the first three years. 
The comparison with Fig.~\ref{flavour} shows that the actual value of the spectral index plays some role in determining the allowed regions: the steeper spectrum $\alpha=2.3$ enhances the role of electron neutrinos and 
diminishes the one of muon neutrinos (whose effective area is very small close to the threshold) thus it 
requires to increase the content of muon neutrinos at the source $\xi_\mu^0$ in order to reproduce the observed track-to-shower ratio. For this reason, the agreement of the neutron decay scenario with  
the data worsens for $\alpha=2.3$. However, this kind of  effects it is not yet crucial for the analysis and in particular the 
neutron decay scenario is not yet excluded. In fact the most important conclusion is just that the small number of events presently available, does not allow us yet to exclude any mechanism of production \cite{Palladino:2015zua}. This  remains true also using 
%with $\alpha=2.3$ (the three years best fit of IceCube's HESE events) or 
$\alpha=2.6$, namely the best fit including also low energy events. 
%\footnote{Anyway it does not seem justified to consider only one spectral index when analyzing low energy events, because background and signal are mixed. }. 

%A novelty of the present analysis is that we show, for the first time, the impact of uncertainty on oscillation parameters in the triangle of flavor, both with Montecarlo approach and Gaussian approach: 
The confidence levels are in reasonable accordance with those of  IceCube data analysis~\cite{Aartsen:2015ivb} (see again figure~\ref{flavour}) and with those of \cite{Palladino:2015zua}.  The uncertainties due to the oscillation parameters 
have been presented  in~\cite{Palladino:2015zua} in a different manner, but the results are in excellent agreement.\footnote{A recent work \cite{salvado} appeared after the present one 
also shows the uncertainties due to oscillations in the flavor triangle. The Monte Carlo procedure is used and the results coincide with our ones.} 
%that this figure shows %for the first time (boh! ce lo mettiamo?
%the impact that the uncertainties in parameters of oscillation have on final flavour ratio.

\begin{figure}[t]
\includegraphics[scale=0.45]{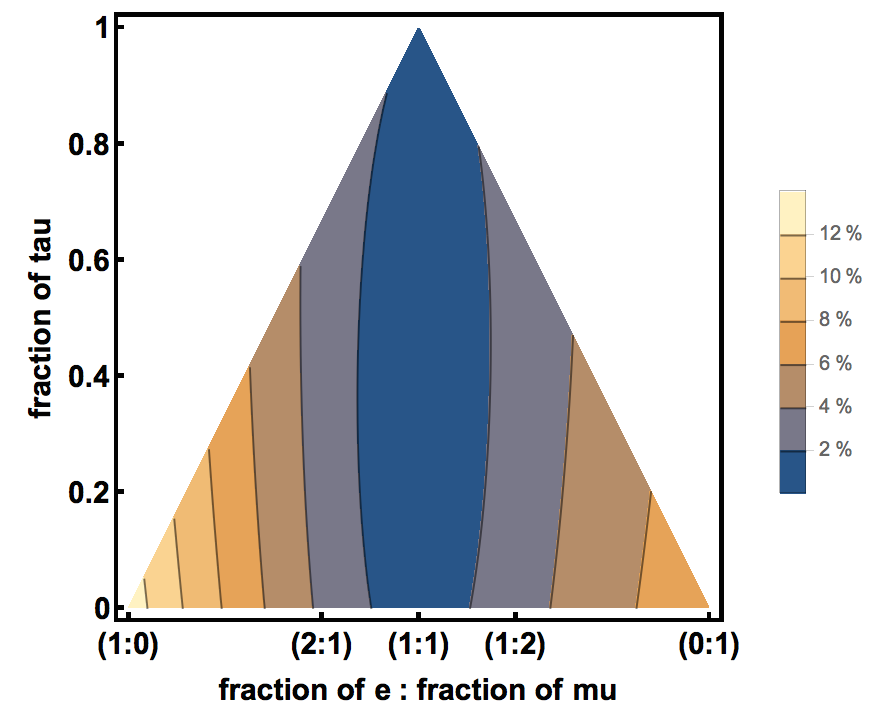}
\centering
\caption{\em \small Gaussian errors on the fraction of $\nu_\mu$  on Earth at 1 sigma, as a function of the neutrino fractions  at the source.}
\label{errmutau}
\end{figure}

\subsection{Errors on flavor ratio} %1 applicazione

Let us discuss further the point of the errors.
The flavor ratios given in Eqs.~\ref{fracmu} depend linearly by $P_0$, $P_1$ and $P_2$.
Therefore, it is straightforward and quite easy to evaluate the Gaussian errors on  these quantities. Let us focus on $\xi_\mu$.

From the formula for 
$\xi_\mu$, again Eq.~(\ref{fracmu}), we see 
that the term linear in $P_1$ becomes very small when 
$\nu_e$ and $\nu_{\mu}$ are about equal (e.g.\ charm mesons),
and Fig.~\ref{errmutau} confirms that  that this type of mechanisms gives very small errors on the flavor ratio measured at Earth. Indeed, an initial composition of 1:1:1 would not be modified, or in other words, the error would be just null). On the contrary, the mechanisms that produce only $\nu_e$ (neutrons decay) or only $\nu_\mu$ (damped muons) give the biggest error, about 10$\%$ on final flavor ratio. The pion decay, that is the most  plausible  mechanism, is between the two extreme situations; the error on the muon fraction $\xi_\mu$ is about~3$\%$.

Let us remark that despite the relatively large uncertainties on $\theta_{23}$ and $\delta$
oscillations, the uncertainties on $\xi_\mu$ are small also in the worse scenario, 
namely, the neutron decay scenario.

These results are in good agreement
 with those of \cite{nl5}, see in particular Fig.~4 there.
Moreover, we note  that 
the expectations from the pion decay  mechanism agree quite well with the results of the analysis of the existing data; see again  Fig.~\ref{flavour}.

\subsection{Glashow resonance}
With the formalism of this paper it is easy to write analytical expressions for some interesting signal including the effect of three flavor oscillations. We analyze the case Glashow resonance  \cite{glash0}, i.e.\ the production of $W^-$ starting from electron antineutrino due to the process,
\begin{equation}
\overline{\nu}_e + e^- \rightarrow W^- \nonumber
\end{equation}
The process is possible when the antineutrino with energy greater than 6.3 PeV collides with an electron at rest.

\begin{figure}[t]
\centering
\includegraphics[scale=0.5]{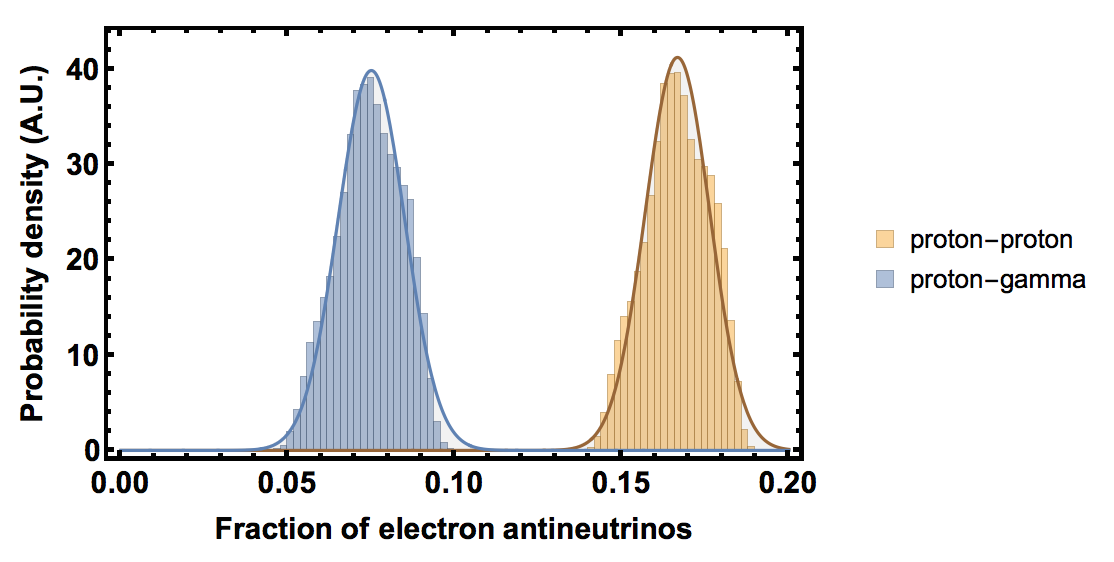}
\caption{\em \small Fraction of $\overline{\nu}_e$ due to $pp$ or $p\gamma$ interaction. The distribution obtained by a Montecarlo extraction compares well with the Gaussian distributions obtained from Table~\ref{tab1} (continuous lines).}
\label{glashow}
\end{figure}

% \cite{glash0}, \cite{glash1} and \cite{glash2}. 

An interesting point for us is that different astrophysical mechanisms produce different fraction of $\overline{\nu}_e$, as already noted in
\cite{piton} and further discussed in \cite{nl5,glash,glash1,glash2}. From the $pp$ interaction, e.g.\ \cite{ppaha}, we can obtain all the type of pions, instead from $p\gamma$ interaction, e.g.\ \cite{pgammaaha}, we obtain mostly $\pi^+$ and $\pi^0$. After the decays the flavor ratio for $pp$ are approximatively equal for neutrinos and antineutrinos,
\begin{equation}
\xi_\nu^0=(1,2,0)/6 \ , \ 
\xi_{\overline{\nu}}^0=(1,2,0)/6
\end{equation}
while for $p \gamma$, the neutrino and antineutrino channels contain
a different number of particles and lead to different flavors,
\begin{equation}
\xi_\nu^0=(1,1,0)/3  \ , \ 
\xi_{\overline{\nu}}^0=(0,1,0)/3 
\end{equation}
where we have normalized the two fluxes to a single
particle. In the case of $p \gamma$ interaction $\overline{\nu}_e$ are not produced at the source, but only after the oscillations. The fraction of electron antineutrinos at Earth are given by a linear expression in the parameters $P_0$ and $P_1$.
\begin{eqnarray}
\xi_{\overline{\nu}_e}^{pp} &=& \frac{1}{6} + \frac{1}{3} P_1 \\[1ex]
\xi_{\overline{\nu}_e}^{p \gamma} &=&\frac{1-3P_0}{9}+ \frac{1}{3} P_1
\end{eqnarray}
These two distribution can be obtained with Monte Carlo simulation, using the distributions of mixing angle and CP phase violation. 
At $3\sigma$ (i.e., 99.7\%) we
find that
\begin{eqnarray}
\xi_{\overline{\nu}_e}^{pp} &=& 0.167\pm 0.029 \\[1ex]
\xi_{\overline{\nu}_e}^{p \gamma} &=&0.075\pm 0.029
\end{eqnarray}
where the  uncertainty is due to $P_1$.
These ranges  compare well with the maximum and minimum values given in Fig.~6 of \cite{nl5}.  
Note that we are using the newest measurements 
of the oscillation parameters and in particular we 
have included the effect of  $\theta_{13}$, 
 that is now measured and known to be non-zero.

%The main feature is that  
 %the two cases of $pp$ or $p\gamma$ sources 
 %are well distinguished 
 %despite the uncertainties on the oscillation parameters.
 If the total flux of neutrinos is the same in both cases, 
this implies that the signal from the Glashow resonance is about two times weaker in the $p \gamma$ case than in the $pp$ case.  Thus, oscillations give a final flavor ratio that is significantly different in the case of $p \gamma$ production mechanism.
The results are further illustrated in figure \ref{glashow}.  
Note that, also for this application, the very small differences with the result of the Gaussian approximation, evident from Fig.~\ref{glashow}, justify the use of a linear approach.

 If we  compare the two cases  assuming to have the same cosmic ray flux and the same amount of collisions, as done in~\cite{piton}, the difference between the two cases is even more dramatic. In fact, the $p\gamma$ mechanism leads to neutrinos of lower energies than those from the $pp$ mechanism, and thus  the chances of observing a neutrino signal from the Glashow resonance
 decreases.

%4-12
%14-20

\section{Summary and discussion}
The results of IceCube have greatly increased the interest in an accurate  description of propagation of cosmic neutrinos, accounting in particular for the minimal hypothesis of three flavor oscillation in vacuum. In this work we have discussed what is the natural choice of the parameters to describe vacuum  oscillations of cosmic neutrinos.

We have shown that 
the parameterization of neutrino oscillations in vacuum introduced in \cite{nl45} 
can be promoted to an exact one  
(without any need of performing expansions, approximations or of making references to  case of tribimaximal neutrino mixing) 
and can be argued to be the natural parameterization for the discussion of  oscillations of cosmic neutrinos, see Sect.~\ref{motta}. 
We have obtained new and exact expressions for the three relevant parameters in terms of the standard mixing parameters, see Sect.~\ref{cx2}.

In Sect.~\ref{appli}, we have illustrated the usefulness of the natural parameters $P_0,P_1,P_2$ given in Eq.~\ref{our} by discussing the expectations on the neutrino flavor ratios and their errors.
We have also analyzed the expectations on the intensity of the signal due Glashow resonance, that depends on the mechanism of neutrino  production. 
We have included   the effect of the uncertainties on oscillation
in the flavor triangle, comparing the predictions with 
the results of the analysis of the flavor of cosmic neutrinos
seen by IceCube, finding results in good agreement with
the previous literature, \cite{Palladino:2015zua} and 
\cite{Aartsen:2015ivb}.
We have confirmed \cite{nl5} that the $pp$ and $p\gamma$ production  mechanism lead to significantly different predictions for the intensity of the Glashow resonance. 

With improved data and analyses of IceCube data, it will be 
more and more important to include the effect of the various 
theoretical uncertainties, including those due to parameters of 
neutrino oscillations. 
Here, we have shown that the expectations obtained with a Gaussian treatment of the natural parameters $P_0,P_1,P_2$ 
are very similar to those obtained in more complete descriptions of three flavor oscillations baseed on~\cite{Palladino:2015zua}. 
We have proved that the Gaussian treatment, particularly  easy to implement, is quite adequate for the present needs.

Let us conclude by stressing that the parameters $P_0,P_1$ and $P_2$ can be used also to provide us with an ideally exact and compact description of cosmic neutrino oscillations. E.g., the distributions of these parameters obtained with MonteCarlo methods, described above and illustrated in the flavor triangles of Figs.~\ref{flavour} and \ref{flavour2},  go beyond the Gaussian approximation and it is formally accurate. 

Future global analyses of the data on three flavor neutrino oscillation will be able to derive which are the precise  distributions of the natural parameters of the cosmic neutrino oscillations and their correlations, simply because these parameters are functions of the conventional mixing angles and CP-violating phase. In view of the above discussion, it will be particularly useful to have precise  distributions of the parameters $P_1$ and $P_2$.

\paragraph{Acknowledgments}
We are grateful to F.~Aharonian, G.~Pagliaroli and F.L.~Villante for useful discussions and collaboration on related matter.  
We thank an anonymous referee of EPJC for pointing out to us the first reference on Glashow resonance and cosmic neutrino oscillations;
F.V.\ gladly acknowledges a discussion on the same topic with Klas Hultqvist at the Neutrino Telescope conference at Venice (March 2015).
%\newpage

%\newpage
\appendix
\section{Allowed regions in the Gaussian approximation\label{aa}}

Let us consider  the two dimensional Gaussian likelihood,
\begin{equation}
\mathcal{L}=
\frac{\exp\left[ -\frac{1}{2} (\vec{v}-\langle\vec{v}\rangle)^t\  {\Sigma^{-2}}\ (\vec{v}-\langle\vec{v}\rangle)\right]}{2 \pi \sqrt{\mbox{det}(\Sigma^2)}}
\mbox{ where  }
	\vec{v}=\left(\begin{array}{c} x \\ y
	\end{array}\right)
	, \ 
	%\vec{\mu}=\left(\begin{array}{c} \mu_x \\ \mu_y
	%\end{array}\right)
	%, \ 
\Sigma^2= 
\left(\begin{array}{cc} \sigma_x^2 &\sigma^2  \\[0.5ex]
\sigma^2 & \sigma_y^2
\end{array}\right)
\end{equation}
A  confidence level ($0<C.L.<1$) defines the allowed region 
$\mathcal{L}>\mathcal{L}_{\mbox{\tiny max}}\ (1-C.L.)$. Its contour is an ellipse that can be obtained  from the following parametric expression, 
\begin{equation}
\left(\!\!\begin{array}{c} {x}(\varphi) \\ 
{y}(\varphi)
	\end{array}\!\!\right)= \left(\!\!\begin{array}{c} \langle{x}\rangle \\ 
\langle{y}\rangle
	\end{array}\!\!\right)+
\sqrt{-2\log(1-C.L.)}
\left(\begin{array}{cc} \cos\theta &\sin\theta  \\ 
  -\sin\theta  & \cos\theta
\end{array}\right)
\left(\begin{array}{c} \sigma_+\ \cos\varphi \\ \sigma_-\ \sin\varphi
\end{array}\right)
\end{equation}
where $\varphi=[0,2\pi]$ and where we defined,
\begin{equation}
\theta=\frac{1}{2}\ \mbox{arctan}\left[ \frac{2 \sigma^2}{
\sigma_y^2- \sigma_x^2}\right]
\ , \ 
\sigma_\pm^2=\frac{2 (\sigma_x^2\sigma_y^2-\sigma^4 ) }{
\sigma_x^2+ \sigma_y^2\pm (\sigma^2_y-\sigma^2_x)
\sqrt{ 1+\tan^2 2 \theta }}
\end{equation}
In the flavor triangle, we have known 
linear combinations of $P_0,P_1,P_2$,
\begin{equation}
x=(\xi_\mu-\xi_e)/\sqrt{3}\equiv x_0 + x_i P_i \mbox{ and }y=\xi_\tau
\equiv y_0 + y_i P_i
\end{equation}
{}From~Tab.~\ref{tab1}, one evaluates 
$\langle x\rangle=x_0+x_i \langle P_i\rangle$,
$\sigma^2_x=x_i^2\, \delta P_i^2$, 
$\sigma^2=x_i y_i\, \delta P_i^2$, etc.\

%\newpage

\end{document}